\documentclass[10pt,aps,prb,twocolumn,superscriptaddress,balancelastpage]{revtex4-1}

\usepackage[pdftex]{graphicx}
\usepackage[colorlinks=true]{hyperref}
\usepackage{subfigure}
\usepackage{bm,amsmath,amssymb}

\newcommand{\ket}[1]{|#1\rangle}

\begin{document}
\title{Prediction of a Weyl Semimetal in Hg$_{1-x-y}$Cd$_x$Mn$_y$Te}
\author{Daniel Bulmash}
\affiliation{Department of Physics, Stanford University, Stanford, California 94305-4045, USA}
\author{Chao-Xing Liu}
\affiliation{Department of Physics, The Pennsylvania State University, University Park, Pennsylvania 16802-6300}
\author{Xiao-Liang Qi}
\affiliation{Department of Physics, Stanford University, Stanford, California 94305-4045, USA}
\date{\today}

\begin{abstract}
We study strained Hg$_{1-x-y}$Cd$_x$Mn$_y$Te in a magnetic field using a $\bm{k}\cdot\bm{p}$ model and predict that the system is a Weyl semimetal with two nodes in an experimentally reasonable region of the phase diagram. We also predict two signatures of the Weyl semimetal phase which arise from tunability of the Weyl node splitting. First, we find that the Hall conductivity is proportional to the average Mn ion spin and thus is strongly temperature dependent. Second, we find an unusual magnetic field angle dependence of the Hall conductivity; in particular, we predict a peak in $\sigma_{xy}$ as a function of field angle in the $xz$-plane and a finite $\sigma_{yz}$ as the $x$-component of the field goes to 0.

\end{abstract}

\maketitle

%\section{Introduction}

\noindent{\bf Introduction.}-- Since the connection of the Chern number to the quantum Hall effect\cite{thouless1982}, topology has become an increasingly important ingredient in classifying phases of matter. Despite all early examples of topological states being insulators, it has recently been shown that gapless materials can also have topologically nontrivial properties, with the primary example being the Weyl semimetal (WSM)\cite{murakami2007,Wan2011,burkov2011topological,hosurReview}.

A WSM is a 3D topological phase characterized by pairs of nondegenerate linear touchings of the bulk bands, called Weyl nodes. Such touchings are hedgehogs in momentum space with quantized topological charge\cite{volovik2009universe}. Weyl nodes always come in pairs of opposite charge due to the Nielsen-Ninomiya theorem\cite{Nielsen_Ninomiya}, and they are topologically protected. In the simple case of two nodes, this protection can be seen by considering a 2D slice of the Brillouin zone (BZ) which is perpendicular to the vector between the Weyl nodes; the Chern number of each low-energy band changes by the charge of the node when this slice crosses a Weyl node\cite{Yang_Lu_Ran_2011}. This implies that Weyl nodes can only be gapped out by annihilating two nodes of opposite charge at the same point in momentum space if charge conservation and translation symmetry are present. Moreover, the nonzero Chern number between the nodes implies the existence of surface states with Fermi arcs, i.e. a Fermi surface that does not form a closed curve but instead terminates at the projection of the Weyl nodes onto the surface momentum space\cite{Wan2011}.

There have so far been several proposals for creating a WSM; candidate structures include pyrochlore iridates\cite{Wan2011}, topological insulator/normal insulator heterostructures\cite{Burkov_multilayer,balents2012WSM}, and HgCr$_2$Se$_4$\cite{xu2011chern}. Two related materials are Na$_3$Bi\cite{WangNa3Bi2012} and Cd$_3$As$_2$\cite{wang2013three}, which are Dirac semimetals, i.e. their Weyl nodes are degenerate.
However, experimental confirmation of these predictions has been limited. DC resistivity measurements of Y$_2$Ir$_2$O$_7$\cite{yanagishima}
are in good agreement with theoretical models\cite{Hosur2012}, hinting at a WSM phase, and there is possible evidence for Weyl physics in Bi$_{0.97}$Sb$_{0.03}$ in a magnetic field\cite{Kim2013}.

In this paper, we predict that in experimentally feasible parameter regimes, strained Hg$_{1-x-y}$Cd$_x$Mn$_y$Te in an external magnetic field is a WSM with one pair of nodes. In order to form a WSM, a "parent" 3D Dirac cone is typically required for each pair of nodes. Upon breaking either inversion or time reversal ($\mathcal{T}$) symmetry, the Dirac cone can split into Weyl nodes separated in momentum space.  In our model, the parent Dirac cone appears when the 3D topological insulator (TI) HgTe\cite{Brune2011,dai2008helical,fuKane2007} is Cd-doped to the TI-normal insulator phase transition. Further doping by Mn breaks $\mathcal{T}$ in the presence of an external magnetic field via exchange between the paramagnetic Mn ions and the $sp$ conduction electrons, producing a WSM.

We emphasize that our model is particularly simple in that it contains the minimum number of Weyl nodes and appears in a material which is experimentally well-controlled. Another advantage is that in our realization, tuning the magnetization of the Mn ions adjusts the splitting of the Weyl points. We demonstrate that this tunability leads to two signatures of the WSM phase which could be used to distingush it from other semimetallic phases. First, for compressive strain in the $z$-direction and $x$-direction magnetic field, the Hall conductivity $\sigma_{yz}$ is proportional to the Mn magnetization, which is strongly temperature dependent. Second, for the same strain, $\sigma_{xy}$ is non-monotonic in magnetic field angle above the $xz$-plane.

%\section{Model}
\noindent{\bf Model.}--Our starting point is the six-band $\bm{k}\cdot \bm{p}$ model for Hg$_{1-x-y}$Cd$_x$Mn$_y$Te with exchange coupling and diagonal strain\cite{novik2005band}. The six bands in this effective model are the $\Gamma_6$ and $\Gamma_8$ bands; we neglect the $\Gamma_7$ bands since they are split from the $\Gamma_6$ and $\Gamma_8$ bands by the spin-orbit scale $\Delta \sim 1$ eV, which is much larger than all other energy scales in the problem. The $\bm{k}\cdot \bm{p}$ approximation breaks down when $kP + \hbar^2k^2/2m$ (with $P$ the Kane matrix element) is of order the gap between the included and neglected bands at $k=0$; this gap is of order 1 eV, which leads to $ka \sim 0.5$, far larger than any momentum that will be considered in this work. This Hamiltonian should therefore be valid for any momentum we consider. The Hamiltonian is also valid at temperatures smaller than this 1 eV gap, i.e. at $\lesssim 10^3$ K.

We perform all calculations within the virtual crystal approximation (VCA). This requires that $y \ll 1$ (an assumption that we also use to neglect the effect of Mn dopants on the band structure parameters) and that the Cd dopants do not order. In principle we should assume $x$ is small as well, but the VCA has yielded excellent agreement with experiment even for quantum wells with $x=0.7$\cite{novik2005band}, so we assume the VCA for all $x$. All lattice parameters are linearly interpolated in $x$ except for the zero-field gap at $y=0$. We use a quadratic interpolation of the last parameter from the literature\cite{laurenti1990temperature}.

In general, the Zeeman and Landau splittings are far smaller than the exchange interaction; for $B \approx 1$ T, the energy scale of the exchange interaction is $\sim (0.8$ eV$)y$, whereas the Zeeman and Landau splitting energy scale $\hbar e B/m$ is about $0.1$ meV. We therefore neglect orbital effects and check the consistency of this assumption later.

\begin{figure}
\includegraphics[width=8cm]{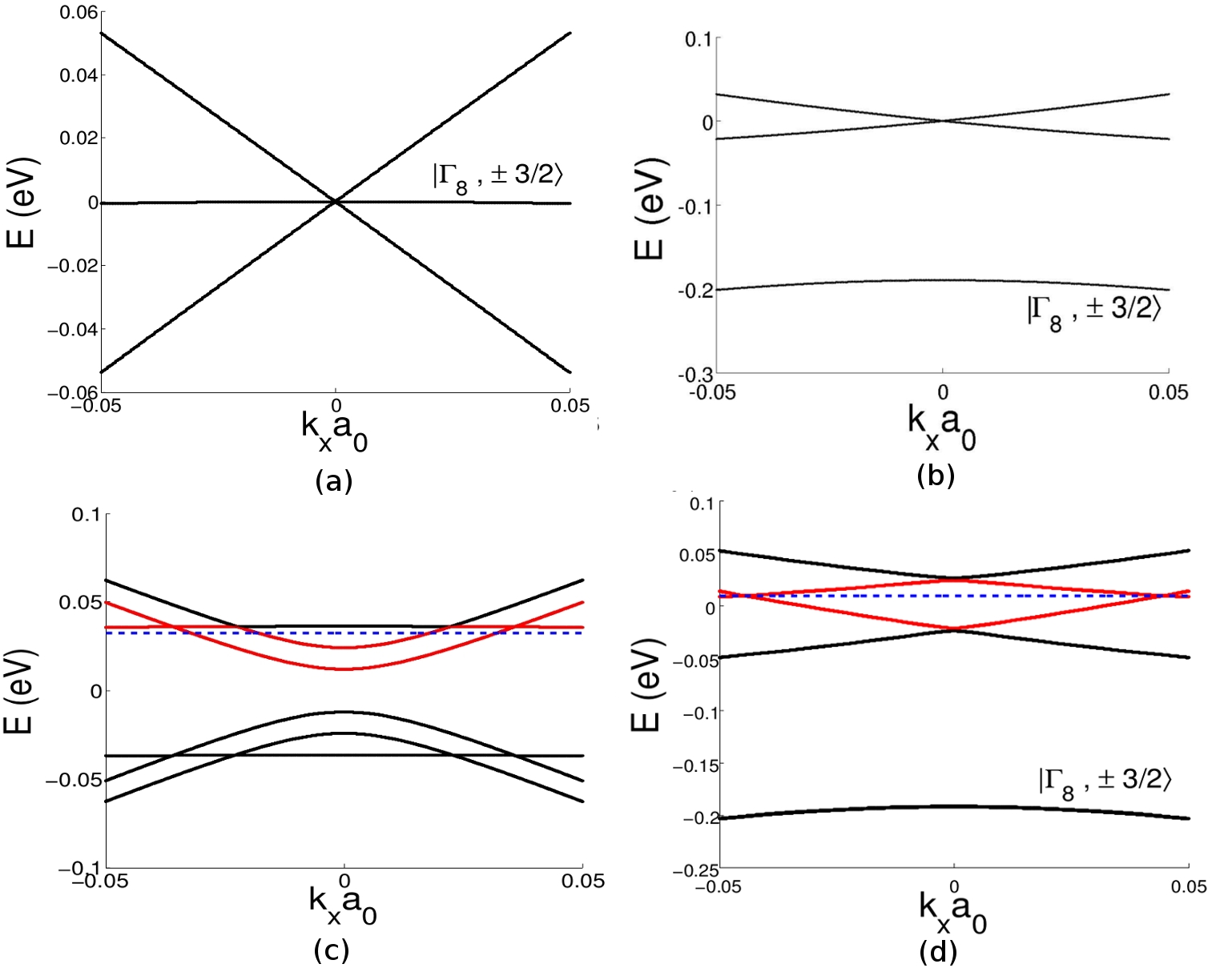}
\caption{Effect of strain and exchange on the continuum band structure near the $\Gamma$ point. The label $\ket{\Gamma_8, \pm 3/2}$ refers to the character at $k=0$. Parameters: $k_y = k_z = 0$, $x$ tuned to set $E_g + E_{str}=0$ (see discussion after Equation \ref{eqn:Hamiltonian} in the text). ``Exchange" means $y=0.05$, $B_x = -5$ T, ``Strain" means $\epsilon_{zz} = -0.04$. (a): No strain, no exchange. (b) Strain only. (c) Exchange only; Weyl bands are shown in red and the Fermi energy is the dashed blue line. (d) Strain and exchange, same colors as (c); the chemical potential crosses the $k \neq 0$ degeneracies.}
\label{fig:bandStructure}
\end{figure}

At stoichiometry, four of the six bands are filled. At zero strain, the $m_J = \pm 3/2$ bands add substantial complication to the band structure, as shown in Fig. \ref{fig:bandStructure}. With compressive $z$-direction strain and/or tensile $x$- or $y$-direction strains, the $m_J = \pm 3/2$ bands decrease in energy. These bands can be neglected at large enough strain, as in Figs. \ref{fig:bandStructure}b and \ref{fig:bandStructure}d, leaving a four-band model at half-filling. We use the basis of band-edge Bloch functions $\lbrace \ket{\Gamma_6, m_J = +1/2},\ket{\Gamma_6, m_J = -1/2}, \ket{\Gamma_8,m_J =+1/2},\ket{\Gamma_8,m_J = -1/2}\rbrace$ for the remaining bands.
We will consider the effects of decreased strain later. We use as a basis for the anticommuting Dirac matrices in the Euclidean metric
$(\Gamma_0, \Gamma_1, \Gamma_2,\Gamma_3, \Gamma_5) = (\tau_z \otimes 1, -\tau_y \otimes \sigma_y,-\tau_y \otimes \sigma_x,\tau_x \otimes 1, \tau_y \otimes \sigma_z)$
where $\tau$ and $\sigma$ act on the orbital and angular momentum degrees of freedom respectively, and let $\Gamma_{ab} = i/2\left[\Gamma_a,\Gamma_b\right]$. Assuming by rotational symmetry that $B_y = 0$, the continuum model can be written (see Supplemental Material\footnote{Supplemental Material can be found at \hyperref[http://link.aps.org/doi/10.1103/PhysRevB.89.081106]{http://link.aps.org/doi/10.1103/PhysRevB.89.081106}})
\begin{align}
H = \sum_{j=1}^3&\left(v_j k_j \Gamma_j + c_jk_j^2 \Gamma_0\right) + \left(E_g + E_{str}\right) \Gamma_0 \nonumber\\
&+ a_x \Gamma_{23} + a_z \Gamma_{12} + b_x  \Gamma_{15} + b_z  \Gamma_{35}
\label{eqn:Hamiltonian}
\end{align}
where $a_{x,z}$ and $b_{x,z}$ are proportional to both the average Mn spin $\langle S_{x,z}\rangle$ in the $x$ and $z$ directions respectively and the Mn doping $y$.
The important tunable parameters are:  $\langle S_j \rangle$, the zero-field, zero-strain band gap $E_g$, and the strain contribution $E_{str}$ to the 4-band Hamiltonian (see Supplemental Material for their explicit forms). The prefactors $v_j$ and $c_j$, as well as the proportionality constants for $a_{x,z}$ and $b_{x,z}$, are phenomenological constants from the $\bm{k}\cdot \bm{p}$ model; some depend on $x$ and strain, but not in a way which qualitatively affects the physics. Due to the particular values for this material, it turns out that $|b_x| \ll |a_x|$, so we neglect $b_x$.

To get the parent Dirac cone, we note that while both $E_g$ and $E_{str}$ depend on $x$, only $E_{str}$ depends on the strain. We then choose to tune $x$ so that $E_g + E_{str} = 0$. We assume that the applied strain has the form $\epsilon_{xx} = \epsilon_{yy} = -\epsilon_{zz}/(2C_{12}/C_{11})$, with the deformation potentials $C_{12}$ and $C_{11}$ taken from the literature\cite{alper1967elastic}. This would be the case for strain due to the lattice mismatch between HgTe and the CdTe substrate, for example. In this case, the constraints $E_g+E_{str}=0$ and $0 \leq x \leq 1$ can be satisfied for any $0\leq \epsilon_{xx} \lesssim 0.8$; the very large upper bound on the strain means that this constraint is easily satisfied in the range of validity of our model. We defer discussion of what happens when $E_g + E_{str}\neq 0$ until later, but note that the topological protection of the Weyl nodes removes the need for perfect fine-tuning.

At low energy, this model is easily diagonalized:
\begin{eqnarray}
E = \pm &\left\lbrace a_x^2 + a_z^2 + b_z^2 + \sum_{i=x,y,z} (v_i k_i)^2 \pm 2 \left[a_z^2b_z^2 + \right. \right. \nonumber\\
&\left. \left.(a_x v_x k_x + a_z v_z k_z)^2 + b_z^2 (v_y^2 k_y^2 + v_x^2 k_x^2)\right]^{1/2} \right\rbrace^{1/2}
\label{eqn:eigenvalues}
\end{eqnarray}
We assume for now that $B_z = 0$ so that $a_z = b_z = 0$. There are exactly three gapless points, appearing at $\bm{k}=0$ and at $\pm k_x^{\ast} = \pm |a_x|/v_x$, $k_y=k_z=0$. The $\bm{k}=0$ degeneracy is not topological and can be lifted by adding infinitesimal $B_z$ or coupling to the $\ket{\Gamma_8,\pm 3/2}$ bands, so we ignore this degeneracy. The low-energy dispersion about the other two gapless points is clearly linear. To confirm that this is indeed a WSM with two nodes, we can consider our Hamiltonian to be a set of 2D systems in the $y$ and $z$ directions, parametrized by $k_x$. This Hamiltonian will be gapped at all points except for $k_x = \pm k^{\ast}$ and at the accidental degeneracy at $k=0$. The Chern number of these quasi-2D Hamiltonians can be computed numerically\cite{fukui2005chern} with the standard lattice regularization $k_i \rightarrow \sin(k_i a_0)/a_0$, $k_i^2 \rightarrow 2(1-\cos(k_i a_0))/a_0^2$ (with $a_0$ the lattice constant). Indeed, for $|k_x| > k_x^{\ast}$, the Chern number of the low-energy bands is zero, but the Chern number becomes $\pm 1$ (depending on the band) for $|k_x| < k_x^{\ast}$. This change in Chern number, a hallmark of the WSM, indicates that the degeneracy at the Weyl points is topologically protected so long as the Weyl points remain separated in momentum space\cite{Wan2011}.

This change in Chern number also implies a quantized Hall conductance $\sigma_{yz} = (2k_x^{\ast}a_0/2\pi)e^2/h$, where $a_0$ is the lattice constant\cite{Yang_Lu_Ran_2011}. That is, the Hall conductance is $e^2/h$ times the fraction of the 2D systems which have a Chern number $\pm 1$ band.

\begin{figure}
  \subfigure{
  \includegraphics[width=4cm]{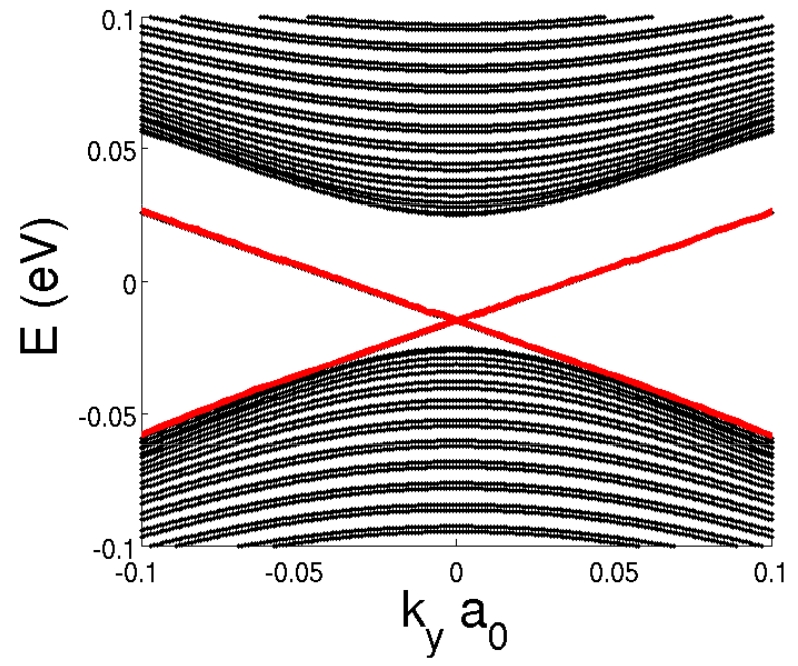}
	 \label{fig:EdgeStates}}
  \subfigure{
  \includegraphics[width=4cm]{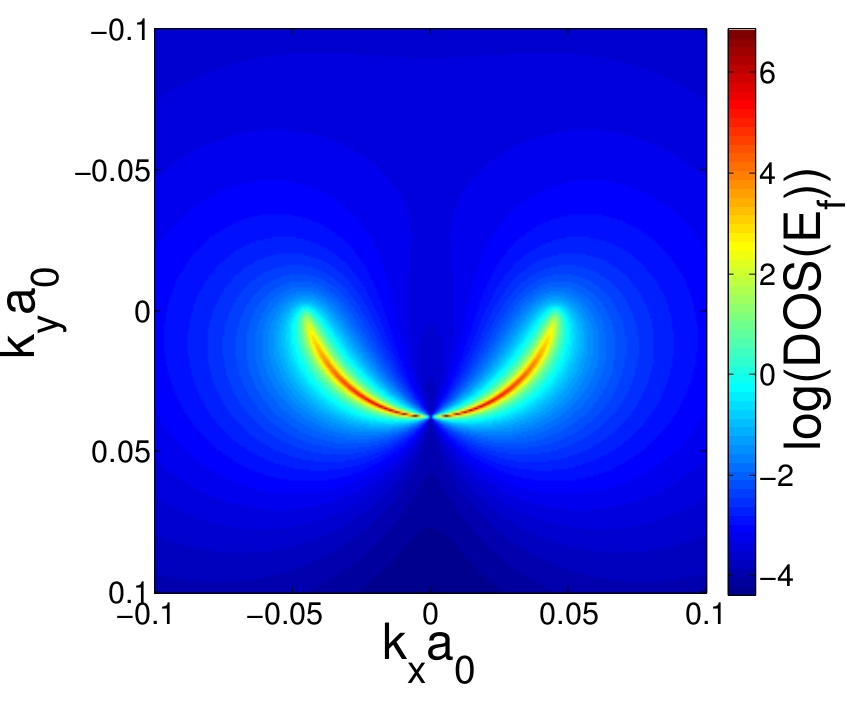}
 \label{fig:fermiArcs}
 }
 \caption{(a): Edge state spectrum in the four-band model for a finite $z$ direction at $k_x = 0$ in a strip geometry. Edge states are colored red. Parameters used: 400 sites in the $z$-direction, $|B_x| = 5$ T, $y=0.05$, critical $x$ neglecting strain. (b): Surface density of states at the Fermi level at the same parameters as in (a) but in a semi-infinite geometry.}
 \end{figure}

To further confirm the WSM phase, we investigate the surface states. We assume that there are edges in the $z$ direction and numerically diagonalize the lattice-regularized version of Equation \ref{eqn:Hamiltonian}. Results are shown in Fig. \ref{fig:EdgeStates}. The edge states clearly exist at small $k_x$, though there is also a finite size gap at the Weyl nodes. We also calculated the surface density of states using a Green's function method\cite{dai2008helical} for a semi-infinite configuration. As expected, the resulting Fermi arc, shown in Fig. \ref{fig:fermiArcs}, smoothly merges into the Weyl points at $k_x = \pm k^{\ast}$.

We now relax our conditions on the strain. First suppose that we allow $E_g + E_{str} = \Delta E \neq 0$. So long as $\Delta E$ is smaller than the exchange energy scale (about 100 meV for $y \sim 1$\%), the Weyl nodes will not annihilate. At fixed strain, $\Delta E \lesssim 100$ meV means that $x$ can vary from optimal by about 2\%. At fixed doping, the strain tolerated varies strongly on the direction in which it is applied; for a strain like the HgTe/CdTe lattice mismatch, the Weyl points tolerate a strain past the validity of our model (the model allows $\Delta \epsilon_{xx} \sim 20$\%), while only a $\sim 1$\% uniaxial strain in the $z$ direction is allowed.

Next consider the effect of allowing small strain, but maintain the constraint $E_g + E_{str} = 0$ by changing $x$. As the strain decreases, we find numerically that the Weyl points stay approximately fixed in momentum. Assuming stoichiometry (so that exactly four bands are full), the Hall conductivity is therefore unchanged until the filling of the individual bands changes. This occurs when the Fermi energy falls below the Weyl points. The system also becomes fully metallic at this point, as in Fig. \ref{fig:bandStructure}c.

\begin{figure}
\subfigure{\includegraphics[width=4cm]{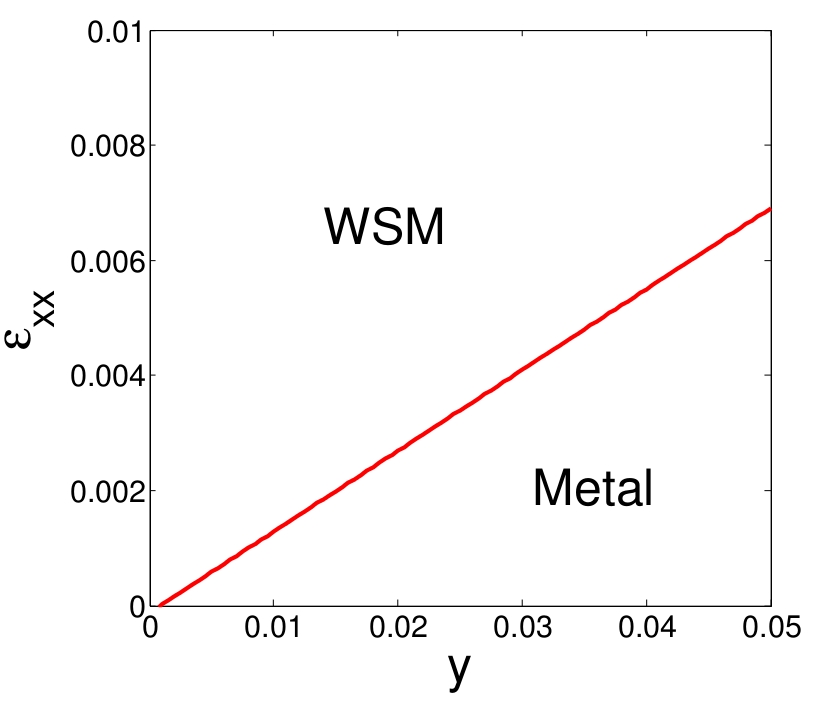}\label{fig:phaseDiagram}}\subfigure{\includegraphics[width=4cm]{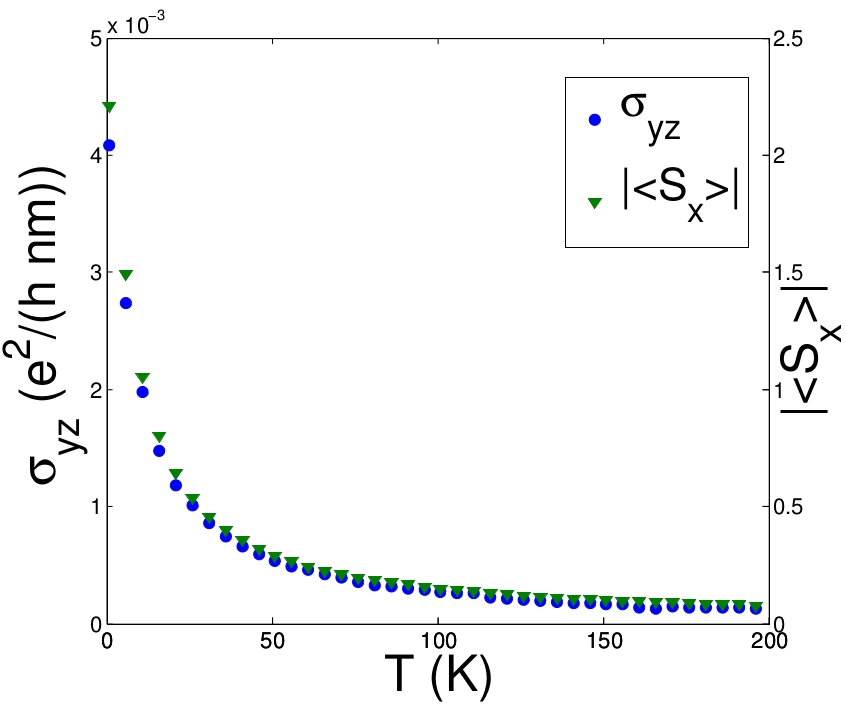}\label{fig:conductivity}}
\caption{(a) Phase diagram at $B=5$ T with $x$ tuned so that $E_g + E_{str} = 0$. The slightly negative intercept of the boundary line is interpreted as numerical error. (b) Hall conductivity $\sigma_{yz}$ and average Mn spin $\langle S_x \rangle$ as a function of temperature in the four-band model at $B_x = 4$ T, $y=0.01$	. The lattice constant has been set to $1$ nm for simplicity.}
%\label{fig:phaseDiagram}
\end{figure}

The loss of the quantized zero-temperature Hall conductivity can be used to define a boundary between the WSM and fully metallic phases. As can be seen from the resulting phase diagram in Fig. \ref{fig:phaseDiagram}, a WSM appears in a reasonable region of parameter space; a field of 5 T (which fully saturates the Mn spins), strain of tenths of a percent, and $y$ of a few percent are all within the realm of experiments. In particular, HgTe on a CdTe substrate has a lattice mismatch of $0.3\%$, which is enough to get a WSM up to $y \approx 0.02$ at the optimal $x=0.16$ for this strain. However, larger strain is preferred; although the WSM phase requires less strain at smaller $y$, the Weyl node splitting and thus all relevant signatures increase with $y$. In particular, $\sigma_{yz} = (2k^{\ast}a/2\pi)(e^2/h)$, which has a maximum value of $0.30y(e^2/h)$. Hence larger strain is ideal to allow a larger $y$, but small values of $y$  also suffice.

We also need to check our assumptions that Zeeman splittings and Landau level formation are not important, assuming that $y$ is large enough that the exchange interaction is large compared to the characteristic energy scales for these effects. The Zeeman splitting has the same form as the exchange term\cite{novik2005band}, but is $y$-independent; in this regime the Zeeman term may therefore certainly be neglected. Next, note that the Landau level splitting does not affect the anomalous Hall effect; the reason is that the Landau levels disperse in $k_x$. The only place that the bands will change filling due to Landau level formation is near the Weyl points, in particular for the $k$ for which the gap is smaller than the cyclotron frequency $v_F \sqrt{eB/\hbar}$ (i.e. they differ from a Weyl point by at most the inverse cyclotron length). For a field of $1$ T, this leads to $\Delta k a_0 \sim 0.02$, or a total momentum space volume $\sim 4\pi(\Delta k a)^3/3 \approx 3 \times 10^{-5}$. However, the anomalous Hall effect is due to the entire region $|k_x| \leq k^{\ast}$, with a momentum space volume $\sim (2\pi)^2k^{\ast}a_0 \sim 2$ for the $k^{\ast}a_0 = 0.05$ from Fig. \ref{fig:bandStructure}d. Since a negligible amount of this latter volume is unaffected, the Hall response should be negligibly affected. Finally, it is possible that in very clean samples Shubnikov-de-Haas oscillations could be observed if the scattering time is larger than the inverse cyclotron frequency $\omega_c^{-1}$. Near the Weyl points, $\hbar \omega_c = \hbar v_F\sqrt{eB\hbar} \sim 2$ meV for $B \sim 1$ T (here we estimated $v_F \sim 10^5$ m/s directly from the band structure).

%\section{Experimental Signatures}
%\label{section:expts}

\noindent{\bf Experimental signatures.}--Since the Mn ions in our model are paramagnetic, a magnetic field is required to realize the WSM. This means that observing the Fermi arcs via angle-resolved photoemission spectroscopy (ARPES) is not feasible, but it allows the Weyl point splitting to be tuned by the temperature and magnetic field. We will now show that this tunability leads to two major signatures of the WSM phase in this material which could reasonably be seen in experiments. The system is assumed to have large enough strain that the four-band model is valid.

First, we consider the temperature dependence of the Hall conductivity $\sigma_{yz}$. The Weyl point splitting $k^{\ast}$ fully determines the intrinsic contribution to the Hall conductivity $\sigma_{yz}$. Increasing the temperature at fixed magnetic field will disorder the Mn spins, decreasing the effect of exchange coupling. In particular, $k^{\ast}$ is proportional to the average dopant spin $\langle \bm{S}\rangle$. Empirically\cite{novik2005band},
\begin{equation}
\langle S_i \rangle = -S_0 B_{5/2}\left(\frac{5g_{Mn}\mu_B B_i}{2k_B(T+T_0)}\right)
\end{equation}
where $B_i$ is the $i$th component of the magnetic field, $B_{5/2}$ is the spin-$5/2$ Brillouin function, $g_{Mn} = 2$ is the $g$ factor of Mn, and the effective spin $S_0$ and effective temperature $T_0$ account for antiferromagnetism between Mn ions. Hence we expect that, barring other temperature-related contributions, $\sigma_{yz}$ will have the same temperature dependence as $\langle S_x \rangle$. The temperature scale for this decrease is approximately $20$ K for magnetic fields of order Teslas. To check this result, we numerically calculated $\sigma_{yz}(T)$ using the Kubo formula in the clean four-band model and plot the results in Fig. \ref{fig:conductivity}.

%\begin{figure}
%\includegraphics[width=5cm]{conductivity.jpg}
%\caption{Hall conductivity $\sigma_{yz}$ and average Mn spin $\langle S_x \rangle$ as a function of temperature in the four-band model at $B_x = 4$ T, $y=0.01$	. The lattice constant has been set to $1$ nm for simplicity.}
%\label{fig:conductivity}
%\end{figure}

The conductivity follows the average Mn spin, indicating that there are no other significant contributions in this limit at zero chemical potential. We further claim that the Hall conductivity is independent of small changes in the chemical potential\cite{Yang_Lu_Ran_2011}. In particular, very close to the Weyl nodes, the low-energy Hamiltonian $H = \chi (\bm{k} - \bm{k}_0)\cdot \bm{\sigma}$ is invariant under the antiunitary operation which takes $\Delta \bm{k} = \bm{k} - \bm{k}_0$ to $-\Delta\bm{k}$ and acts like time reversal on spin. The Berry curvature is odd under this pseudo-time-reversal operation, so any state and its pseudo-time-reversed partner have equal and opposite contributions to the Berry curvature (and thus the Hall conductivity). Thus the region of momentum space well-described by this low energy Hamiltonian has zero net contribution to the Hall conductivity, so changing the chemical potential has no effect on the Hall conductivity so long as it remains at low energy.

\begin{figure}
\subfigure{\includegraphics[width=4cm]{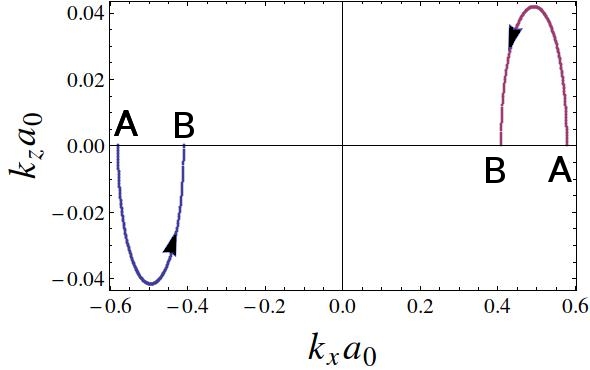}
\label{fig:WeylNodePath}}\subfigure{\includegraphics[width=4cm]{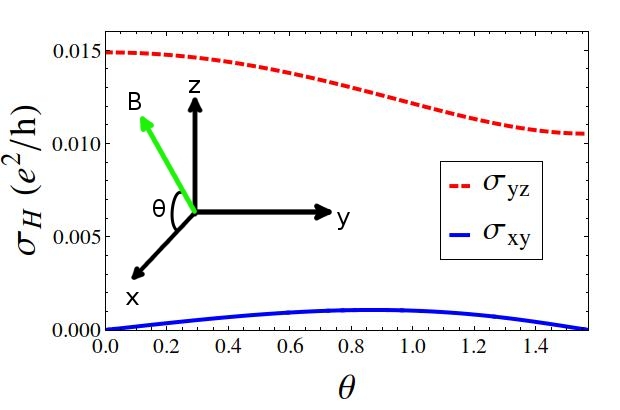}
\label{fig:HallAngledB}}
\caption{(a): Path of the Weyl nodes in momentum space as $\theta$ changes. A and B correspond to $\theta=0$ and $\theta = \pi/2$ respectively. (b): Hall conductivities $\sigma_{xy}$, $\sigma_{yz}$ as a function of magnetic field angle $\theta$. }
\end{figure}

As a second probe, we consider the Hall conductivity as a function of magnetic field angle $\theta$ above the $xy$-plane. We can see from Equation \ref{eqn:eigenvalues} that when $B_z \neq 0$, all degeneracies must occur at zero energy. The resulting equation for degeneracies becomes quartic and its analytical solution becomes unwieldy, but we can find numerically that the Weyl nodes follow a path in the $xz$-plane, shown in Fig. \ref{fig:WeylNodePath}. ``Slicing" the BZ in the $k_x$ and $k_z$ planes yields $\sigma_{yz}$ and $\sigma_{xy}$ resepectively, as the conductivity is the fraction of the BZ which lies between the slices with the Weyl nodes. The results are shown in Fig. \ref{fig:HallAngledB}. The most striking feature is that $\sigma_{xy}$ is non-monotonic in $\theta$. Also, $\sigma_{yz}$ does not go to zero as $\theta$ goes to $\pi/2$ because the Weyl node splitting is not along the field.

There is a complication to these results; at $\theta = \pi/2$ the system forms a ring node that overlaps with the Weyl nodes. This ring node is not topologically interesting because, like typical ring nodes\cite{burkov2011topological}, it is unstable, but the additional gap closings as $\theta \rightarrow \pi/2$ increase the relevance of finite temperature. Fortunately, for values of $y$ of a few percent, the $\theta = 0$ gap at the eventual location of the ring node is of order tens of Kelvin, so at liquid He temperatures finite temperature effects should not interfere with observation of $\sigma_{xy}$ until well past its peak.

So far we have only considered a clean system. Because our system is gapless, at any finite disorder concentration we expect side-jump and skew-scattering contributions to the Hall conductivity\cite{Nagaosa2010}.
However, none of these contributions should show the non-monotonicity in $\sigma_{xy}$ that appears in the tilted-field case, and non-magnetic impurities should contribute to $\sigma_{yz}$ with a very different temperature dependence (likely on a different temperature scale as well). Moreover, disorder-driven contributions should be small; free carriers are usually needed for such contributions, and the density of states at the Fermi level vanishes in the ideal WSM case, so we may be in a regime analogous to a band insulator. It is, however, difficult to quantitatively treat this regime, as typical perturbation expansions in $1/k_Fl_{mfp}$ with $l_{mfp}$ the mean free path break down near a Dirac point.

%\begin{acknowledgments}
\noindent{\bf Acknowledgements.}--We would like to thank Pavan Hosur and Laurens W. Molenkamp for helpful discussions. DB is supported in part by the National Science Foundation Graduate Research Fellowship under Grant No. DGE-114747. XLQ is supported by the Defense Advanced
Research Projects Agency Microsystems Technology
Office, MesoDynamic Architecture Program (MESO)
through the contract number N66001-11-1-4105.
%\end{acknowledgments}

\bibliographystyle{apsrev4-1}
\bibliography{HgTeProjectBib}

%Merlin.mbs v4.21 2009-07-09.
\begin{thebibliography}{10}%
\makeatletter
\providecommand \@ifxundefined [1]{%
 \ifx #1\undefined \expandafter \@firstoftwo
 \else \expandafter \@secondoftwo
\fi
}%
\providecommand \@ifnum [1]{%
 \ifnum #1\expandafter \@firstoftwo
 \else \expandafter \@secondoftwo
\fi
}%
\providecommand \enquote [1]{``#1''}%
\providecommand \bibnamefont  [1]{#1}%
\providecommand \bibfnamefont [1]{#1}%
\providecommand \citenamefont [1]{#1}%
\providecommand\href[0]{\@sanitize\@href}%
\providecommand\@href[1]{\endgroup\@@startlink{#1}\endgroup\@@href}%
\providecommand\@@href[1]{#1\@@endlink}%
\providecommand \@sanitize [0]{\begingroup\catcode`\&12\catcode`\#12\relax}%
\@ifxundefined \pdfoutput {\@firstoftwo}{%
 \@ifnum{\z@=\pdfoutput}{\@firstoftwo}{\@secondoftwo}%
}{%
 \providecommand\@@startlink[1]{\leavevmode\special{html:<a href="#1">}}%
 \providecommand\@@endlink[0]{\special{html:</a>}}%
}{%
 \providecommand\@@startlink[1]{%
  \leavevmode
  \pdfstartlink
   attr{/Border[0 0 1 ]/H/I/C[0 1 1]}%
   user{/Subtype/Link/A<</Type/Action/S/URI/URI(#1)>>}%
  \relax
 }%
 \providecommand\@@endlink[0]{\pdfendlink}%
}%
\providecommand \url  [0]{\begingroup\@sanitize \@url }%
\providecommand \@url [1]{\endgroup\@href {#1}{\urlprefix}}%
\providecommand \urlprefix [0]{URL }%
\providecommand \Eprint[0]{\href }%
\@ifxundefined \urlstyle {%
  \providecommand \doi [1]{doi:\discretionary{}{}{}#1}%
}{%
  \providecommand \doi [0]{doi:\discretionary{}{}{}\begingroup
  \urlstyle{rm}\Url }%
}%
\providecommand \doibase [0]{http://dx.doi.org/}%
\providecommand \Doi[1]{\href{\doibase#1}}%
\providecommand \bibAnnote [3]{%
  \BibitemShut{#1}%
  \begin{quotation}\noindent
    \textsc{Key:}\ #2\\\textsc{Annotation:}\ #3%
  \end{quotation}%
}%
\providecommand \bibAnnoteFile [2]{%
  \IfFileExists{#2}{\bibAnnote {#1} {#2} {\input{#2}}}{}%
}%
\providecommand \typeout [0]{\immediate \write \m@ne }%
\providecommand \selectlanguage [0]{\@gobble}%
\providecommand \bibinfo [0]{\@secondoftwo}%
\providecommand \bibfield [0]{\@secondoftwo}%
\providecommand \translation [1]{[#1]}%
\providecommand \BibitemOpen[0]{}%
\providecommand \bibitemStop [0]{}%
\providecommand \bibitemNoStop [0]{.\EOS\space}%
\providecommand \EOS [0]{\spacefactor3000\relax}%
\providecommand \BibitemShut [1]{\csname bibitem#1\endcsname}%
%</preamble>
\bibitem{thouless1982}%
  \BibitemOpen
  \bibfield{author}{%
  \bibinfo {author} {\bibfnamefont{D.~J.}\ \bibnamefont{Thouless}}, \bibinfo
  {author} {\bibfnamefont{M.}~\bibnamefont{Kohmoto}}, \bibinfo {author}
  {\bibfnamefont{M.~P.}\ \bibnamefont{Nightingale}},\ and\ \bibinfo {author}
  {\bibfnamefont{M.}~\bibnamefont{den Nijs}},\ }%
  \bibfield{journal}{%
  \Doi{10.1103/PhysRevLett.49.405}{\bibinfo {journal} {Phys. Rev. Lett.}}\ }%
  \textbf{\bibinfo {volume} {49}},\ \bibinfo {pages} {405} (\bibinfo {year}
  {1982})%
  \bibAnnoteFile{NoStop}{thouless1982}%
\bibitem{murakami2007}%
  \BibitemOpen
  \bibfield{author}{%
  \bibinfo {author} {\bibfnamefont{S.}~\bibnamefont{Murakami}},\ }%
  \bibfield{journal}{%
  \Doi{10.1088/1367-2630/9/9/356}{\bibinfo {journal} {New J. Phys.}}\ }%
  \textbf{\bibinfo {volume} {9}},\ \bibinfo {pages} {356} (\bibinfo {year}
  {2007})%
  \bibAnnoteFile{NoStop}{murakami2007}%
\bibitem{Wan2011}%
  \BibitemOpen
  \bibfield{author}{%
  \bibinfo {author} {\bibfnamefont{X.}~\bibnamefont{Wan}}, \bibinfo {author}
  {\bibfnamefont{A.~M.}\ \bibnamefont{Turner}}, \bibinfo {author}
  {\bibfnamefont{A.}~\bibnamefont{Vishwanath}},\ and\ \bibinfo {author}
  {\bibfnamefont{S.~Y.}\ \bibnamefont{Savrasov}},\ }%
  \bibfield{journal}{%
  \Doi{10.1103/PhysRevB.83.205101}{\bibinfo {journal} {Phys. Rev. B}}\ }%
  \textbf{\bibinfo {volume} {83}},\ \bibinfo {pages} {205101} (\bibinfo {year}
  {2011})%
  \bibAnnoteFile{NoStop}{Wan2011}%
\bibitem{burkov2011topological}%
  \BibitemOpen
  \bibfield{author}{%
  \bibinfo {author} {\bibfnamefont{A.~A.}\ \bibnamefont{Burkov}}, \bibinfo
  {author} {\bibfnamefont{M.~D.}\ \bibnamefont{Hook}},\ and\ \bibinfo {author}
  {\bibfnamefont{L.}~\bibnamefont{Balents}},\ }%
  \bibfield{journal}{%
  \Doi{10.1103/PhysRevB.84.235126}{\bibinfo {journal} {Phys. Rev. B}}\ }%
  \textbf{\bibinfo {volume} {84}},\ \bibinfo {pages} {235126} (\bibinfo {year}
  {2011})%
  \bibAnnoteFile{NoStop}{burkov2011topological}%
\bibitem{hosurReview}%
  \BibitemOpen
  \bibfield{author}{%
  \bibinfo {author} {\bibfnamefont{P.}~\bibnamefont{Hosur}}\ and\ \bibinfo
  {author} {\bibfnamefont{X.}~\bibnamefont{Qi}},\ }%
  \bibfield{journal}{%
  \bibinfo {journal} {e-print}\ }%
  \Eprint{http://arxiv.org/abs/cond-mat/1309.4464}{cond-mat/1309.4464}%
  \bibAnnoteFile{NoStop}{hosurReview}%
\bibitem{volovik2009universe}%
  \BibitemOpen
  \bibfield{author}{%
  \bibinfo {author} {\bibfnamefont{G.~E.}\ \bibnamefont{Volovik}},\ }%
  \emph{\bibinfo {title} {The universe in a helium droplet}}\ (\bibinfo
  {publisher} {Oxford University Press},\ \bibinfo {year} {2009})%
  \bibAnnoteFile{NoStop}{volovik2009universe}%
\bibitem{Nielsen_Ninomiya}%
  \BibitemOpen
  \bibfield{author}{%
  \bibinfo {author} {\bibfnamefont{H.}~\bibnamefont{Nielsen}}\ and\ \bibinfo
  {author} {\bibfnamefont{M.}~\bibnamefont{Ninomiya}},\ }%
  \bibfield{journal}{%
  \Doi{10.1016/0370-2693(83)91529-0}{\bibinfo {journal} {Phys. Lett. B}}\ }%
  \textbf{\bibinfo {volume} {105}},\ \bibinfo {pages} {219} (\bibinfo {year}
  {1981})%
  \bibAnnoteFile{NoStop}{Nielsen_Ninomiya}%
\bibitem{Yang_Lu_Ran_2011}%
  \BibitemOpen
  \bibfield{author}{%
  \bibinfo {author} {\bibfnamefont{K.-Y.}\ \bibnamefont{Yang}}, \bibinfo
  {author} {\bibfnamefont{Y.-M.}\ \bibnamefont{Lu}},\ and\ \bibinfo {author}
  {\bibfnamefont{Y.}~\bibnamefont{Ran}},\ }%
  \bibfield{journal}{%
  \Doi{10.1103/PhysRevB.84.075129}{\bibinfo {journal} {Phys. Rev. B}}\ }%
  \textbf{\bibinfo {volume} {84}},\ \bibinfo {pages} {075129} (\bibinfo {year}
  {2011})%
  \bibAnnoteFile{NoStop}{Yang_Lu_Ran_2011}%
\bibitem{Burkov_multilayer}%
  \BibitemOpen
  \bibfield{author}{%
  \bibinfo {author} {\bibfnamefont{A.~A.}\ \bibnamefont{Burkov}}\ and\ \bibinfo
  {author} {\bibfnamefont{L.}~\bibnamefont{Balents}},\ }%
  \bibfield{journal}{%
  \Doi{10.1103/PhysRevLett.107.127205}{\bibinfo {journal} {Phys. Rev. Lett.}}\
  }%
  \textbf{\bibinfo {volume} {107}},\ \bibinfo {pages} {127205} (\bibinfo {year}
  {2011})%
  \bibAnnoteFile{NoStop}{Burkov_multilayer}%
\bibitem{balents2012WSM}%
  \BibitemOpen
  \bibfield{author}{%
  \bibinfo {author} {\bibfnamefont{G.~B.}\ \bibnamefont{Hal\'asz}}\ and\
  \bibinfo {author} {\bibfnamefont{L.}~\bibnamefont{Balents}},\ }%
  \bibfield{journal}{%
  \Doi{10.1103/PhysRevB.85.035103}{\bibinfo {journal} {Phys. Rev. B}}\ }%
  \textbf{\bibinfo {volume} {85}},\ \bibinfo {pages} {035103} (\bibinfo {year}
  {2012})%
  \bibAnnoteFile{NoStop}{balents2012WSM}%
\bibitem{xu2011chern}%
  \BibitemOpen
  \bibfield{author}{%
  \bibinfo {author} {\bibfnamefont{G.}~\bibnamefont{Xu}}, \bibinfo {author}
  {\bibfnamefont{H.}~\bibnamefont{Weng}}, \bibinfo {author}
  {\bibfnamefont{Z.}~\bibnamefont{Wang}}, \bibinfo {author}
  {\bibfnamefont{X.}~\bibnamefont{Dai}},\ and\ \bibinfo {author}
  {\bibfnamefont{Z.}~\bibnamefont{Fang}},\ }%
  \bibfield{journal}{%
  \Doi{10.1103/PhysRevLett.107.186806}{\bibinfo {journal} {Phys. Rev. Lett.}}\
  }%
  \textbf{\bibinfo {volume} {107}},\ \bibinfo {pages} {186806} (\bibinfo {year}
  {2011})%
  \bibAnnoteFile{NoStop}{xu2011chern}%
\bibitem{WangNa3Bi2012}%
  \BibitemOpen
  \bibfield{author}{%
  \bibinfo {author} {\bibfnamefont{Z.}~\bibnamefont{Wang}}, \bibinfo {author}
  {\bibfnamefont{Y.}~\bibnamefont{Sun}}, \bibinfo {author}
  {\bibfnamefont{X.-Q.}\ \bibnamefont{Chen}}, \bibinfo {author}
  {\bibfnamefont{C.}~\bibnamefont{Franchini}}, \bibinfo {author}
  {\bibfnamefont{G.}~\bibnamefont{Xu}}, \bibinfo {author}
  {\bibfnamefont{H.}~\bibnamefont{Weng}}, \bibinfo {author}
  {\bibfnamefont{X.}~\bibnamefont{Dai}},\ and\ \bibinfo {author}
  {\bibfnamefont{Z.}~\bibnamefont{Fang}},\ }%
  \bibfield{journal}{%
  \Doi{10.1103/PhysRevB.85.195320}{\bibinfo {journal} {Phys. Rev. B}}\ }%
  \textbf{\bibinfo {volume} {85}},\ \bibinfo {pages} {195320} (\bibinfo {year}
  {2012})%
  \bibAnnoteFile{NoStop}{WangNa3Bi2012}%
\bibitem{wang2013three}%
  \BibitemOpen
  \bibfield{author}{%
  \bibinfo {author} {\bibfnamefont{Z.}~\bibnamefont{Wang}}, \bibinfo {author}
  {\bibfnamefont{H.}~\bibnamefont{Weng}}, \bibinfo {author}
  {\bibfnamefont{Q.}~\bibnamefont{Wu}}, \bibinfo {author}
  {\bibfnamefont{X.}~\bibnamefont{Dai}},\ and\ \bibinfo {author}
  {\bibfnamefont{Z.}~\bibnamefont{Fang}},\ }%
  \bibfield{journal}{%
  \bibinfo {journal} {e-print}}%
   (\bibinfo {year} {2013}),\
  \Eprint{http://arxiv.org/abs/arXiv:1305.6780}{arXiv:1305.6780}%
  \bibAnnoteFile{NoStop}{wang2013three}%
\bibitem{yanagishima}%
  \BibitemOpen
  \bibfield{author}{%
  \bibinfo {author} {\bibfnamefont{D.}~\bibnamefont{Yanagishima}}\ and\
  \bibinfo {author} {\bibfnamefont{Y.}~\bibnamefont{Maeno}},\ }%
  \bibfield{journal}{%
  \Doi{10.1143/JPSJ.70.2880}{\bibinfo {journal} {J. Phys. Soc. Jpn.}}\ }%
  \textbf{\bibinfo {volume} {70}},\ \bibinfo {pages} {2880} (\bibinfo {year}
  {2001})%
  \bibAnnoteFile{NoStop}{yanagishima}%
\bibitem{Hosur2012}%
  \BibitemOpen
  \bibfield{author}{%
  \bibinfo {author} {\bibfnamefont{P.}~\bibnamefont{Hosur}}, \bibinfo {author}
  {\bibfnamefont{S.~A.}\ \bibnamefont{Parameswaran}},\ and\ \bibinfo {author}
  {\bibfnamefont{A.}~\bibnamefont{Vishwanath}},\ }%
  \bibfield{journal}{%
  \Doi{10.1103/PhysRevLett.108.046602}{\bibinfo {journal} {Phys. Rev. Lett.}}\
  }%
  \textbf{\bibinfo {volume} {108}},\ \bibinfo {pages} {046602} (\bibinfo {year}
  {2012})%
  \bibAnnoteFile{NoStop}{Hosur2012}%
\bibitem{Kim2013}%
  \BibitemOpen
  \bibfield{author}{%
  \bibinfo {author} {\bibfnamefont{H.-J.}\ \bibnamefont{Kim}}, \bibinfo
  {author} {\bibfnamefont{K.-S.}\ \bibnamefont{Kim}}, \bibinfo {author}
  {\bibfnamefont{J.}~\bibnamefont{Wang}}, \bibinfo {author}
  {\bibfnamefont{M.}~\bibnamefont{Sasaki}}, \bibinfo {author}
  {\bibfnamefont{N.}~\bibnamefont{Satoh}}, \bibinfo {author}
  {\bibfnamefont{A.}~\bibnamefont{Ohnishi}}, \bibinfo {author}
  {\bibfnamefont{M.}~\bibnamefont{Kitaura}}, \bibinfo {author}
  {\bibfnamefont{M.}~\bibnamefont{Yang}},\ and\ \bibinfo {author}
  {\bibfnamefont{L.}~\bibnamefont{Li}},\ }%
  \bibfield{journal}{%
  \bibinfo {journal} {e-print}\ }%
  \Eprint{http://arxiv.org/abs/arXiv:1307.6990v1}{arXiv:1307.6990v1}%
  \bibAnnoteFile{NoStop}{Kim2013}%
\bibitem{Brune2011}%
  \BibitemOpen
  \bibfield{author}{%
  \bibinfo {author} {\bibfnamefont{C.}~\bibnamefont{Br\"une}}, \bibinfo
  {author} {\bibfnamefont{C.~X.}\ \bibnamefont{Liu}}, \bibinfo {author}
  {\bibfnamefont{E.~G.}\ \bibnamefont{Novik}}, \bibinfo {author}
  {\bibfnamefont{E.~M.}\ \bibnamefont{Hankiewicz}}, \bibinfo {author}
  {\bibfnamefont{H.}~\bibnamefont{Buhmann}}, \bibinfo {author}
  {\bibfnamefont{Y.~L.}\ \bibnamefont{Chen}}, \bibinfo {author}
  {\bibfnamefont{X.~L.}\ \bibnamefont{Qi}}, \bibinfo {author}
  {\bibfnamefont{Z.~X.}\ \bibnamefont{Shen}}, \bibinfo {author}
  {\bibfnamefont{S.~C.}\ \bibnamefont{Zhang}},\ and\ \bibinfo {author}
  {\bibfnamefont{L.~W.}\ \bibnamefont{Molenkamp}},\ }%
  \bibfield{journal}{%
  \Doi{10.1103/PhysRevLett.106.126803}{\bibinfo {journal} {Phys. Rev. Lett.}}\
  }%
  \textbf{\bibinfo {volume} {106}},\ \bibinfo {pages} {126803} (\bibinfo {year}
  {2011})%
  \bibAnnoteFile{NoStop}{Brune2011}%
\bibitem{dai2008helical}%
  \BibitemOpen
  \bibfield{author}{%
  \bibinfo {author} {\bibfnamefont{X.}~\bibnamefont{Dai}}, \bibinfo {author}
  {\bibfnamefont{T.~L.}\ \bibnamefont{Hughes}}, \bibinfo {author}
  {\bibfnamefont{X.-L.}\ \bibnamefont{Qi}}, \bibinfo {author}
  {\bibfnamefont{Z.}~\bibnamefont{Fang}},\ and\ \bibinfo {author}
  {\bibfnamefont{S.-C.}\ \bibnamefont{Zhang}},\ }%
  \bibfield{journal}{%
  \Doi{10.1103/PhysRevB.77.125319}{\bibinfo {journal} {Phys. Rev. B}}\ }%
  \textbf{\bibinfo {volume} {77}},\ \bibinfo {pages} {125319} (\bibinfo {year}
  {2008})%
  \bibAnnoteFile{NoStop}{dai2008helical}%
\bibitem{fuKane2007}%
  \BibitemOpen
  \bibfield{author}{%
  \bibinfo {author} {\bibfnamefont{L.}~\bibnamefont{Fu}}\ and\ \bibinfo
  {author} {\bibfnamefont{C.~L.}\ \bibnamefont{Kane}},\ }%
  \bibfield{journal}{%
  \Doi{10.1103/PhysRevB.76.045312}{\bibinfo {journal} {Phys. Rev. B}}\ }%
  \textbf{\bibinfo {volume} {76}},\ \bibinfo {pages} {045302} (\bibinfo {year}
  {2007})%
  \bibAnnoteFile{NoStop}{fuKane2007}%
\bibitem{novik2005band}%
  \BibitemOpen
  \bibfield{author}{%
  \bibinfo {author} {\bibfnamefont{E.~G.}\ \bibnamefont{Novik}}, \bibinfo
  {author} {\bibfnamefont{A.}~\bibnamefont{Pfeuffer-Jeschke}}, \bibinfo
  {author} {\bibfnamefont{T.}~\bibnamefont{Jungwirth}}, \bibinfo {author}
  {\bibfnamefont{V.}~\bibnamefont{Latussek}}, \bibinfo {author}
  {\bibfnamefont{C.~R.}\ \bibnamefont{Becker}}, \bibinfo {author}
  {\bibfnamefont{G.}~\bibnamefont{Landwehr}}, \bibinfo {author}
  {\bibfnamefont{H.}~\bibnamefont{Buhmann}},\ and\ \bibinfo {author}
  {\bibfnamefont{L.~W.}\ \bibnamefont{Molenkamp}},\ }%
  \bibfield{journal}{%
  \Doi{10.1103/PhysRevB.72.035321}{\bibinfo {journal} {Phys. Rev. B}}\ }%
  \textbf{\bibinfo {volume} {72}},\ \bibinfo {pages} {035321} (\bibinfo {year}
  {2005})%
  \bibAnnoteFile{NoStop}{novik2005band}%
\bibitem{laurenti1990temperature}%
  \BibitemOpen
  \bibfield{author}{%
  \bibinfo {author} {\bibfnamefont{J.}~\bibnamefont{Laurenti}}, \bibinfo
  {author} {\bibfnamefont{J.}~\bibnamefont{Camassel}}, \bibinfo {author}
  {\bibfnamefont{A.}~\bibnamefont{Bouhemadou}}, \bibinfo {author}
  {\bibfnamefont{B.}~\bibnamefont{Toulouse}}, \bibinfo {author}
  {\bibfnamefont{R.}~\bibnamefont{Legros}},\ and\ \bibinfo {author}
  {\bibfnamefont{A.}~\bibnamefont{Lusson}},\ }%
  \bibfield{journal}{%
  \Doi{10.1063/1.345119}{\bibinfo {journal} {J. Appl. Phys.}}\ }%
  \textbf{\bibinfo {volume} {67}},\ \bibinfo {pages} {6454} (\bibinfo {year}
  {1990})%
  \bibAnnoteFile{NoStop}{laurenti1990temperature}%
\bibitem{Note1}%
  \BibitemOpen
  \bibinfo {note} {Supplemental Material can be found at \protect \hyperref
  [http://link.aps.org/doi/10.1103/PhysRevB.89.081106]{http://link.aps.org/doi%
/10.1103/PhysRevB.89.081106})}%
  \bibAnnoteFile{NoStop}{Note1}%
\bibitem{alper1967elastic}%
  \BibitemOpen
  \bibfield{author}{%
  \bibinfo {author} {\bibfnamefont{T.}~\bibnamefont{Alper}}\ and\ \bibinfo
  {author} {\bibfnamefont{G.}~\bibnamefont{Saunders}},\ }%
  \bibfield{journal}{%
  \Doi{10.1016/0022-3697(67)90135-7}{\bibinfo {journal} {J. Phys. Chem.
  Solids}}\ }%
  \textbf{\bibinfo {volume} {28}},\ \bibinfo {pages} {1637} (\bibinfo {year}
  {1967})%
  \bibAnnoteFile{NoStop}{alper1967elastic}%
\bibitem{fukui2005chern}%
  \BibitemOpen
  \bibfield{author}{%
  \bibinfo {author} {\bibfnamefont{T.}~\bibnamefont{Fukui}}, \bibinfo {author}
  {\bibfnamefont{Y.}~\bibnamefont{Hatsugai}},\ and\ \bibinfo {author}
  {\bibfnamefont{H.}~\bibnamefont{Suzuki}},\ }%
  \bibfield{journal}{%
  \bibinfo {journal} {e-print}\ }%
  \Eprint{http://arxiv.org/abs/cond-mat/0503172}{cond-mat/0503172}%
  \bibAnnoteFile{NoStop}{fukui2005chern}%
\bibitem{Nagaosa2010}%
  \BibitemOpen
  \bibfield{author}{%
  \bibinfo {author} {\bibfnamefont{N.}~\bibnamefont{Nagaosa}}, \bibinfo
  {author} {\bibfnamefont{J.}~\bibnamefont{Sinova}}, \bibinfo {author}
  {\bibfnamefont{S.}~\bibnamefont{Onoda}}, \bibinfo {author}
  {\bibfnamefont{A.~H.}\ \bibnamefont{MacDonald}},\ and\ \bibinfo {author}
  {\bibfnamefont{N.~P.}\ \bibnamefont{Ong}},\ }%
  \bibfield{journal}{%
  \Doi{10.1103/RevModPhys.82.1539}{\bibinfo {journal} {Rev. Mod. Phys.}}\ }%
  \textbf{\bibinfo {volume} {82}},\ \bibinfo {pages} {1539} (\bibinfo {year}
  {2010})%
  \bibAnnoteFile{NoStop}{Nagaosa2010}%
\end{thebibliography}%

\end{document}